\documentclass[]{aa}
\usepackage{graphicx}
\usepackage{dcolumn}
\usepackage{amsmath}
\usepackage{amssymb}
\usepackage{amscd}
\usepackage{setspace}

\begin{document}
   \title{Trans-ethyl methyl ether in space}
 
   \subtitle{A new look at a complex molecule in selected hot core regions}

   \author{G. W. Fuchs \inst{1,2}, U. Fuchs\inst{3},  T.F. Giesen\inst{3}, 
          F. Wyrowski\inst{4}
         \fnmsep
          }

   \offprints{G. W. Fuchs}

   \institute{Department of Chemistry, UC Berkeley, Berkeley, CA 94720; 
         \and 
              Raymond and Beverly Sackler Laboratory for Astrophysics, Leiden Observatory, \\
              Leiden University, Postbus 9513, 2300 RA, Leiden, The Netherlands \\  
              \email{fuchs@strw.leidenuniv.nl}
         \and
                I. Physikalisches Institut, Universit\"at zu K\"oln,
              Z\"ulpicher Str. 77, D-50937 K\"oln\\
              \email{ufuchs@ph1.uni-koeln.de, giesen@ph1.uni-koeln.de}
         \and
             Max-Planck-Institut f\"ur Radioastronomie, Auf dem H\"ugel 69, 
             D-53121 Bonn \\
             \email{wyrowski@mpifr-bonn.mpg.de}
                      }

   \date{submitted}

   \abstract{
             An extensive search for the complex molecule trans-ethyl methyl ether towards several hot core regions 
             has been performed. Using the IRAM 30m telescope and the SEST 15m 
             we looked at several frequencies where  
             trans-ethyl methyl ether has strong transitions, as well as lines which are particularly sensitive to 
             the physical conditions in which the molecule can be found. 
             We included G34.26,  NGC6334(I), Orion KL, and W51e2 
             which have previously been proven to have a rich chemistry of complex molecules.
             Our observations cannot confirm the tentative Orion KL detection made by \cite{Charnley-2001-SCAPA-57-685} 
             within their stated column density limits, but we confirm 
             the existence of the trans-ethyl methyl ether towards W51e2 with a column density 
             of $2\times10^{14}$ cm$^{-2}$. 
             The dimethyl ether/methanol ratio of 0.6 as well as the newly found ethyl methyl ether/ethanol ratio of 0.13 indicate
             relative high abundances of ethers toward W51e2. Furthermore, the observation of ethyl methyl ether also confirms the 
             importance of ethanol as a grain mantle constituent. 
             We present new upper limits of around 8$\times$10$^{13}$ cm$^{-2}$ for the 
             column densities of the molecule toward Orion KL, G34.26, NGC6334(I) and estimate the column density 
             towards SgrB2(N) to be of the same order. 
             The W51e2 observations are discussed in more detail.                             
   }
 
  \titlerunning{Trans-ethyl methyl ether in space}
  \authorrunning{G. W. Fuchs et al.
              }
  \maketitle

\newcolumntype{.}{D{.}{.}{-1}}
\newcolumntype{d}{D{.}{}{-1}}
\newcolumntype{s}{D{.}{.}{4}}
\newcolumntype{S}{D{.}{.}{7}}
\newcolumntype{g}[1]{D{.}{.}{#1}}

\section{Introduction}

  Various large and complex molecules of prebiotic importance have been found 
  as constituents of dense interstellar clouds but their detection remains difficult and 
  some of them doubtful \cite[]{Snyder-2005-ApJ-619-914}.
  Fortunately, space offers many places with a rich chemistry and hot molecular cores (HMCs) 
  are prime targets for such investigations. However, the spectra towards hot cores often 
  reveal a high line density and thus proper line assignments are difficult and have to be viewed with a critical eye.  
  The ethyl methyl ether (CH$_3$OC$_2$H$_5$) is an asymmetric top molecule with two methyl groups acting as internal rotors.
  Ethyl methyl ether (EME) has two conformers where the trans-EME is lower in energy than the gauche-EME by 550 cm$^{-1}$ 
  \cite[]{Durig-2002-SC-13-1} and thus favored as an astrophysically detectable species.
  We searched for several transitions of trans-EME to confirm the
  tentative detection of this molecule by \cite{Charnley-2001-SCAPA-57-685}. 
  A positive result means the detection of one of the biggest molecules
  in the interstellar medium (ISM). 
   \cite{Charnley-2001-SCAPA-57-685} observed a single line in W51 e1/e2 and Orion KL at 160.1 GHz and another line
  in SgrB2(N) at 79.6 GHz which they assigned to trans-ethyl methyl ether. 
  Based on 
  these isolated measurements,  \cite{Charnley-2001-SCAPA-57-685}
  estimate the column density of EME to be in the range of 10$^{14}$ - 10$^{15}$ cm$^{-2}$ in Sgr B2(N).
  Thus, assuming a hydrogen column density of 10$^{24}$ cm$^{-2}$ gives a fractional abundance of 10$^{-10}$ - 10$^{-9}$. 
  For W51, they estimate EME to have an abundance of a few times 10$^{-10}$.
  Furthermore, they conclude that the emission of EME closely follows that of 
  methanol, as 
  expected if these molecules are chemically linked. 
  \cite{Charnley-1995-ApJ-448-232} suggest the following production mechanism for EME in 
  hot cores where temperatures can rise to 100-300 K. 
  Icy mantles which are thought to contain alcohols (e.g. methanol, ethanol) evaporate 
  from dusty grains, e.g. \cite{Millar-1998-FD-109-15},  
  the alcohols themselves are thought to be produced via surface reactions starting from the 
  precursor CO. 
  Once the mantles evaporate the alcohols can be converted into ethers
  and other species. 
  The production of ethers is thought to happen via ion-molecule
  reactions\footnote{see  \cite{Mackay-1979-CJC-57-1518,Bohme-1979-JACS-101-3724}} (i), 
  methylisation of the protonated species\footnote{see \cite{Karpas-1989-JPC-93-1859}} (ii) 
  and dissociative recombination 
  with electrons (iii):   
  \begin{equation*}
  \begin{CD}
  \text{i)}  \quad \rm C_2H_5OH @>{\rm H_3O^+}>{\rm HCO^+}> \rm C_2H_5OH_2^+ \\
   \text{ii)} \quad \rm C_2H_5OH_2^+  @>>{\rm CH_3OH}> \rm CH_3OC_2H_6^+ \\
   \text{iii)} \quad \rm CH_3OC_2H_6^+ @>{\rm e^-}>> \rm CH_3OC_2H_5
  \end{CD}
  \end{equation*}
  In hot cores saturated species are the dominant constituents.
  For instance, methanol CH$_3$OH is highly
  abundant and dimethyl ether  (CH$_3$)$_2$O and ethanol C$_2$H$_5$OH are seen only
  in hot cores \cite[]{Herbst-2001-CSR-30-168}.
  \cite{Charnley-1995-ApJ-448-232} mention in their work that dimethyl
  ether, diethyl ether (C$_2$H$_5$)$_2$O, and trans-EME
  should be present in detectable quantities within
  cores rich in ethanol and methanol.
  Furthermore, a detection of either EME or diethyl ether would confirm the
  importance of ethanol as a grain mantle constituent.

\section{EME Spectrum and Observations}

\begin{figure*}[t]
   \centering
   \includegraphics[angle=270,width=17cm]{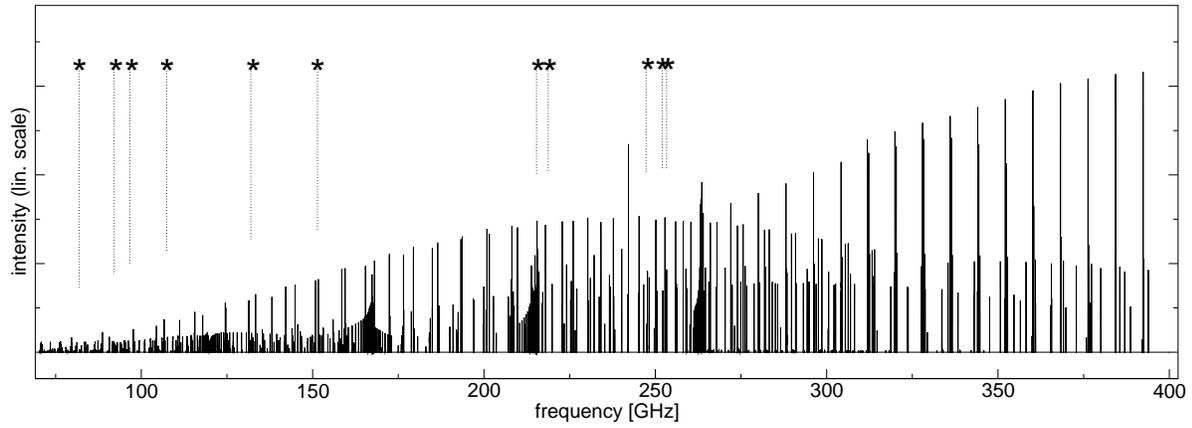}
   \caption{Stick spectrum of EME at a rotational temperature of 130 K which is typical for complex molecules in hot core regions. 
            EME has a dense spectrum with many characteristic features. The  
            frequencies are taken from \cite{Fuchs-2003-ApJSS-144-277}. The ($\star$) indicate our chosen frequency regions 
            for the W51e2 obseravtions. 
            The intense line at 237 GHz was not selected because of a blending line at the same frequency.}
            \label{EME-stick}
    \end{figure*}

  Only two investigations on the laboratory rotational spectrum of EME
  have been reported in the literature.
  The more recent 
  measurements by \cite{Fuchs-2003-ApJSS-144-277} from 55 GHz to 350 GHz also include the previously published data below 40 GHz 
  by  \cite{Hayashi-1975-JMSt-28-147} in their analysis.
  Reliable frequency predictions up to 400 GHz are now available. 
  Since EME consists of two methyl groups designated as internal rotors, the laboratory spectra reveal 
  extra line splittings which are not necessarily seen in space due to larger line widths in the HMC environment. 
  The spectroscopic aspects of ethyl methyl ether in the mm-wave region are described in our earlier publications 
  (see \cite{Fuchs-2003-ApJSS-144-277, U-Fuchs-2003-Cuvillier} and references therein). 
  EME  is an asymmetric top molecule which has weak {\it a}-type transitions and fairly strong {\it b}-type transitions
  above 100 GHz 
  which in theory split into quintets caused by internal torsion. 
  Laboratory measurements can resolve only doublets and quartets
  with splittings mostly between 0.8 and 2 MHz in our examined frequency ranges.
  However, none of the astronomically observed lines is torsionally resolved. 
  Fig.~\ref{EME-stick} shows the complexity of the EME spectrum up to 400~GHz with its characteristic features.

   The IRAM 30m telescope at Pico Veleta, Spain,  has been used during several days in Sept 2002 and July 2003 and 
   the SEST 15m telescope at La Silla, Chile, in March 2003.
   The hot core regions of G34.26, NGC6334I, Orion KL and W51e2 were observed, see Table~\ref{astro-sou}.
   Rotational transitions of EME have been searched  in the 1mm-, 2mm-, and 3mm-range. 
   Depending on the source, between 5 and 11 frequency settings have been used, see
   Table~\ref{obspar} for observational details. Typical frequency resolutions were 0.3, 1 and 1.25 MHz, typical bandwidths 
   between 490 and 1020 MHz and rms noise levels  between 0.01 K at 81 GHz and 0.02 - 0.05 K at 245 GHz.
   The bands were chosen to allow for the observation of CH$_3$OH (a-type), C$_2$H$_5$OH and CH$_3$OCH$_3$
   transitions as well. At least 5 strong lines have been measured from each of these species, 
   except for CH$_3$OH in Orion KL where only three transitions have been observed.
  
   The intensities are measured in $T_{\rm A}^*$ the effective antenna temperature corrected for spillover losses and 
   atmospheric attenuation, 
   and are then converted to $T_{\rm mb}=T_{\rm A}^* / \rho$ using the beam efficiencies $\rho$ of the telescopes at the given frequencies. 
   
   All sources were examined  using long integration times up to 220 minutes on+off time, 
   elevations greater 40$^{\circ}$ and 
   under good atmospheric weather conditions.  


  \section{Source selection}
\label{selection}

    A selection of previous work on the sources is given below. The authors do not intend to give a 
    review of these sources but rather focus on contributions relevant for the EME analysis.
 
  {\bf G34.3+0.15} has been surveyed by \cite{Macdonald-1996-AASS-119-333} and \cite{Thompson-1999-AA-342-809} 
  in the 330-360 GHz region. \cite{Nomura-2004-AA-414-409} examined the chemical structure of this region 
  and gave an overview of the observed species and their abundances 
  along with some comparisons to their model predictions. 
  Single molecule investigations like ethanol have been performed by \cite{Millar-1995-MNRAS-273-25}.
  \cite{Nummelin-1998-AA-337-275} examined CH$_3$CHO and c-C$_2$H$_4$O but also gave abundances for methanol, ethanol 
   and dimethyl ether (DME). 
   \cite{Heaton-1989-AA-213-148} mapped the G34.3 region using the NH$_3$(3,3) emission and determined a size of $3''\times1''.2$.
    The column density of H$_2$ towards G34.26 has been estimated by \cite{Millar-1995-MNRAS-273-25} to be 
    $N_{\rm t}$(H$_2$) = 5.3 $\times$ 10$^{23}$ cm$^{-2}$. \cite{Nummelin-1998-AA-337-275} published a similar result 
    of $N_{\rm t}$(H$_2$) = 3.0 $\times$ 10$^{23}$ cm$^{-2}$ for a 20$''$ source size.
  
  {\bf NGC6334I} (NGC6334F, G351.41+0.64) has been mapped in the continuum by \cite{Sandell-2000-AA-358-242}. 
  \cite{McCutcheon-2000-MNRAS-316-152} took several spectra in the range between 
  334 and 348 GHz using the JCMT. They also mapped the region in CH$_3$OCH$_3$ emission. 
  \cite{Nummelin-1998-AA-337-275} estimated the abundances of ethanol, methanol and DME in 
  NGC6334I and a hydrogen column density of $N_{\rm t}$(H$_2$) = 2.0 $\times$ 10$^{23}$ cm$^{-2}$ within 20$''$. 
  \cite{Cheung-1978-ApJ-226-L149} estimate N$_{H_2}$  $\approx$ 10$^{24}$ cm$^{-2}$. 

  {\bf Orion KL} is a well studied object in the sky.  
  \cite{Turner-1989-ApJSS-70-539,Turner-1991-ApJSS-76-617} 
  performed a line survey between 70 and 115 GHz,
  \cite{Lee-2002-JKAS-35-187} presented line spectra between 159.7 to 164.7 GHz and
  \cite{Sutton-1985-ApJSS-58-341} scanned the spectral region between 215 and 247 GHz.   
  \cite{Blake-1987-ApJ-315-621} studied the 
  chemical composition of OMC-1 in the 230 GHz atmospheric window. They estimate the hydrogen column density  to be 
  N$_{H_2}$ = 1.0 $\times$ 10$^{23}$ cm$^{-2}$ for the hot core.
  The region between 330 to 355 GHz was examined by 
  \cite{Sutton-1991-ApJSS-77-255}, and \cite{Schilke-1997-ApJSS-108-301} performed a survey from  325 GHz to 360 GHz 
  (see also references therein).
  The distribution of some key molecules for the EME analysis has been presented by \cite{Sutton-1995-ApJSS-97-455}.  
  \cite{Caselli-1993-ApJ-408-548} applied a dynamical-chemical model which included  
  abundances of many observed species.   

  The {\bf W51}  molecular cloud is described in \cite{Carpenter-1998-AJ-116-1856}.
  The chemistry of the W51e2 region has been investigated by several groups.   
  For example, \cite{Zhang-1998-ApJ-494-636} has done CH$_3$CN and CS (3-2) observations,  
  \cite{Remijan-2004-ApJ-606-917, Remijan-2002-ApJ-576-264} investigated   CH$_3$CN and CH$_3$COOH,  
   \cite{Liu-2001-ApJ-552-654} looked for HCOOH in this source and \cite{Kuan-2004-ASR-33-31} 
   searched for several molecules of potential prebiotic importance. 
   The hydrogen column density 3.6 $\times$ 10$^{23}$ cm$^{-2}$ \cite[]{Ikeda-2001-ApJ-560-792} is  
   based on C$^{18}$O data from \cite{Schloerb-1987-ApJ-319-426}.

  Also the {\bf SgrB2} region has been investigated by many groups. 
  It is not part of our own observations but since \cite{Charnley-2001-SCAPA-57-685} detected 
  one line at an EME transition frequency we use the published data to estimate the column density of 
  EME in Section~\ref{disc}.    
  A spectral line survey from 70 and 150 GHz has been performed by \cite{Cummins-1986-ApJSS-60-819}, 
  and from 70 to 115 GHz by \cite{Turner-1989-ApJSS-70-539,Turner-1991-ApJSS-76-617}. 
  A further 3mm spectral line survey has been carried out by \cite{Friedel-2004-ApJ-600-234} using the NRAO12m and 
  the BIMA array telescopes. 
  \cite{Nummelin-1998-ApJSS-117-427} extended these observations to frequencies
  between 218 and 263 GHz.  
  Selected molecules, like the oxygen bearing  
  CH$_3$COOH, have been studied by \cite{Remijan-2002-ApJ-576-264}, and NH$_3$ by 
  \cite{Huettenmeister-1993-AA-276-445} with comparisons between several sources 
  including those studied here.  
  \cite{Martin-Pintado-1990-AA-236-193} estimated N$_{H_2}$ = 5.0 $\times$ 10$^{24}$ cm$^{-2}$ to be the hydrogen column density
\footnote{Note, \cite{Charnley-2001-SCAPA-57-685} assume  N$_{H_2}$ = 1.0 $\times$ 10$^{24}$ cm$^{-2}$ }.

  \cite{Ikeda-2001-ApJ-560-792,Ikeda-2002-ApJ-571-560} gave temperatures and abundances of methanol, 
  ethanol and in most cases also for DME for all above mentioned sources. 
  Table~\ref{abundance-tab} summerizes the CH$_3$OH, C$_2$H$_5$OH, CH$_3$OCH$_3$ and C$_2$H$_5$CH$_3$ abundances
  including our latest results.


\section{Data Analysis method}
  \label{analysis}

\begin{figure*}[b]
   \centering
   \includegraphics[angle=270,width=9cm]{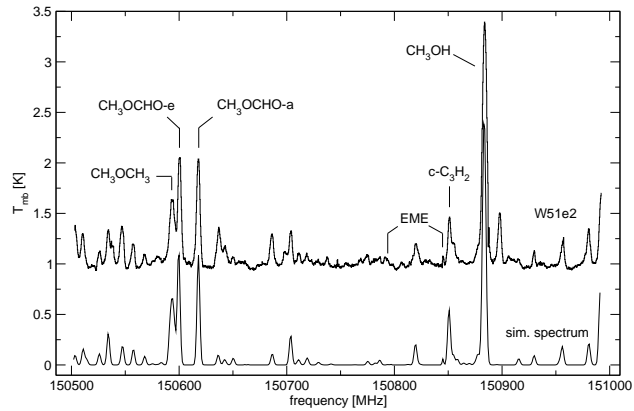}
   \caption{Observed spectrum of W51e2 at 151 GHz (upper) and simulated spectrum using myXCLASS (lower) 
            including C$_2$H$_5$OCH$_3$, C$_2$H$_5$OOCH, CH$_3$CH$_3$CO, a-(CH$_2$OH)$_2$, c-C$_2$H$_4$O,          
            C$_2$H$_5$OH, CH$_3$OCH$_3$, CH$_3$CHO-e, CH$_3$OH, C$_2$H$_5$OH, CH$_3$OCHO-e, and CH$_3$OCHO-a .}
              \label{compare-151GHz-global-W51e2}
    \end{figure*}

  For the analysis of the observed lines we used the  Grenoble Image and Line Data Analysis Software 
  GILDAS\footnote{see http://www.iram.fr/IRAMFR/GILDAS} 
  and in addition the myXCLASS code written by Peter Schilke
  which is backed by a database of astrophysically important molecules,
  including the EME data. 
  myXCLASS can simulate the spectra of many molecules at a time, including their line profiles within the GILDAS environment 
  and has restricted fitting 
  capabilities. 
  Fig.~\ref{compare-151GHz-global-W51e2} shows the observed spectrum of W51e2 at 151 GHz and the synthetic spectrum 
  calculated by myXCLASS assuming optically thin lines where for each molecule the excitation is characterized by a 
  rotational thermal equilibrium temperature. 
  For each source a simulated spectrum has been created which covered all measured frequency bands. 
  This enables an estimate of the abundances and temperatures of the molecules in these sources and can help to reveal 
  blendings of several lines at a given frequency position.
  myXCLASS is specially suited to deal with many convoluting line profiles with different line widths and intensities as 
  can regularly be observed in hot core regions where lines of different molecules blend with each other.    

    The spatial extent of the molecule can influence the value of the inferred column density. This is because 
    the circular Gaussian beam size diameter $\theta_B$ of the telescope varies with frequency $\nu$, i.e. $\theta_B \sim 1/\nu$
    (see telescope beam HPBW in Table~\ref{obspar}).
    If the telescope beam points towards a source  of circular Gaussian type with size $\theta_S$ the resulting convolution 
    beam filling factor $b$ is given by
    $b = \theta_S^2 / (\theta^2_B + \theta_S^2)$, see \cite{Snyder-2005-ApJ-619-914} and the work by 
    \cite{Ulich-1976-ApJSS-30-247}. 
    In the case of $\theta_S \gg \theta_B$ the beam filling factor is close to unity, for  $\theta_S \sim \theta_B$ the 
    beam dilution is 0.5 and 
    for  $\theta_S$ $\ll$  $\theta_B$  it follows $b \approx \theta_S^2/\theta_B^2$. 
    Thus, assuming that the intrinsic intensity distribution $T_{\rm S}$ of a source is Gaussian, a beam filling factor 
    can be introduced to calculate the measured intensity $T_{\rm mb} = b \cdot T_{\rm S}$.       
    HMCs are compact objects with diameters $<10''$. Using the IRAM 30m telescope with HPBW around 30$''$ and 10$''$ 
    at 80 and 250 GHz respectively and assuming a $10''$ source size the beam filling factor is $b$(@80GHz)= 0.1 
    and $b$(@250GHz)=0.5. Hence, an intensity difference 
    of a factor 5 can occur by not correcting for beam dilution. 
    The program myXCLASS has the source size as an input parameter and can thus simulate the effect of the assumed source size 
    on the line intensities. 

    For the EME lines we can expect small optical depths so that the rotational temperature $T_{\rm rot}$ and the total molecular 
    column density $N_{\rm T}$ [cm$^{-2}$] can be determined by a least-squares fit of 
      \begin{equation}
         \log_{10} \left( \frac{3k_B}{8\pi^3}  \frac{\int T_{mb} dv}{\nu\mu^2S} \right) =\log_{10} \left( \frac{N_T}{Q(T_{rot})} \right)
           - \frac{\log_{10}(e)\, E_u}{T_{rot}}
      \end{equation}
    with  $\int T_{mb} dv$ the observed line integral, $\mu^2 S$ the dipole and line strength value derived from 
    \cite{Pickett-1998-JQSRT-60-883} (see their Eq.(2)), $Q$ the partition function, and $E_{\rm u}$ the energy of the upper level. 
    The resulting fit can be plotted in a rotational-temperature-diagram (RTD) or Boltzmann plot.
    For detections at the confusion limit \cite{Snyder-2005-ApJ-619-914} has shown that  Boltzmann plots cannot necessarily 
    be used as independent tools to verify interstellar 
    spectral line identifications. 
    Any false assignment does not automatically reveal itself in the RTD by large displacements 
    from the fitted line and thus can only be avoided by careful line check in catalogs and the literature. 
    As a standard procedure any candidate line of EME has to be checked 
    against competing lines, i.e. the expected or known  intensities of these other lines at the EME frequencies have to be determined.   
    Therfore, a consistent and comprehensive list of molecule abundances in the source is important. Line surveys are invaluable 
    aids in determining the chemical composition of a source.


\section{Results}

\subsection{W51e2}

\begin{figure*}[thH]
   \centering
   \includegraphics[angle=270,width=17cm]{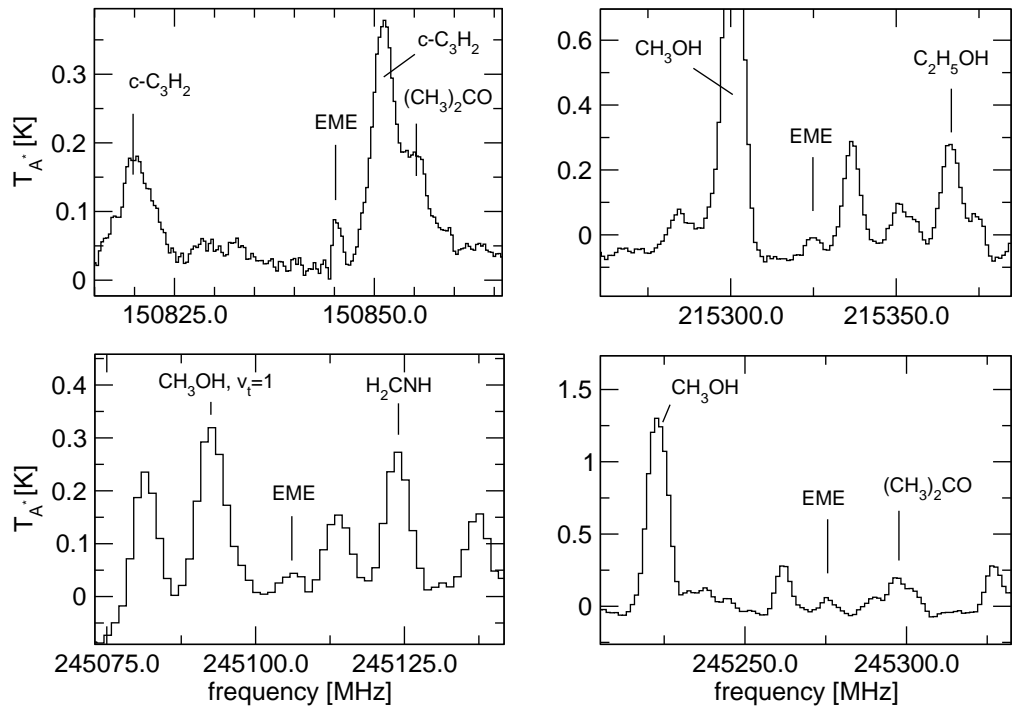}
   \vspace{-0.5cm}\\
   \caption{Spectra recorded towards W51e2. 
             The EME transitions 20$_{0,20}$-19$_{1,19}$ at 150\,845.5 MHz,  
                                 19$_{5,x}$-19$_{4,15}$ at 215\,325.4 MHz , 
                                 16$_{3,14}$-15$_{2,13}$ at 245\,106.4 MHz , and 
                                 17$_{3,15}$-16$_{2,14}$ at 245\,190.7 MHz 
             are indicated. 
               }
              \label{w51-EME}
    \end{figure*}

\begin{figure*}[pth]
   \centering
   \includegraphics[angle=270,width=16cm]{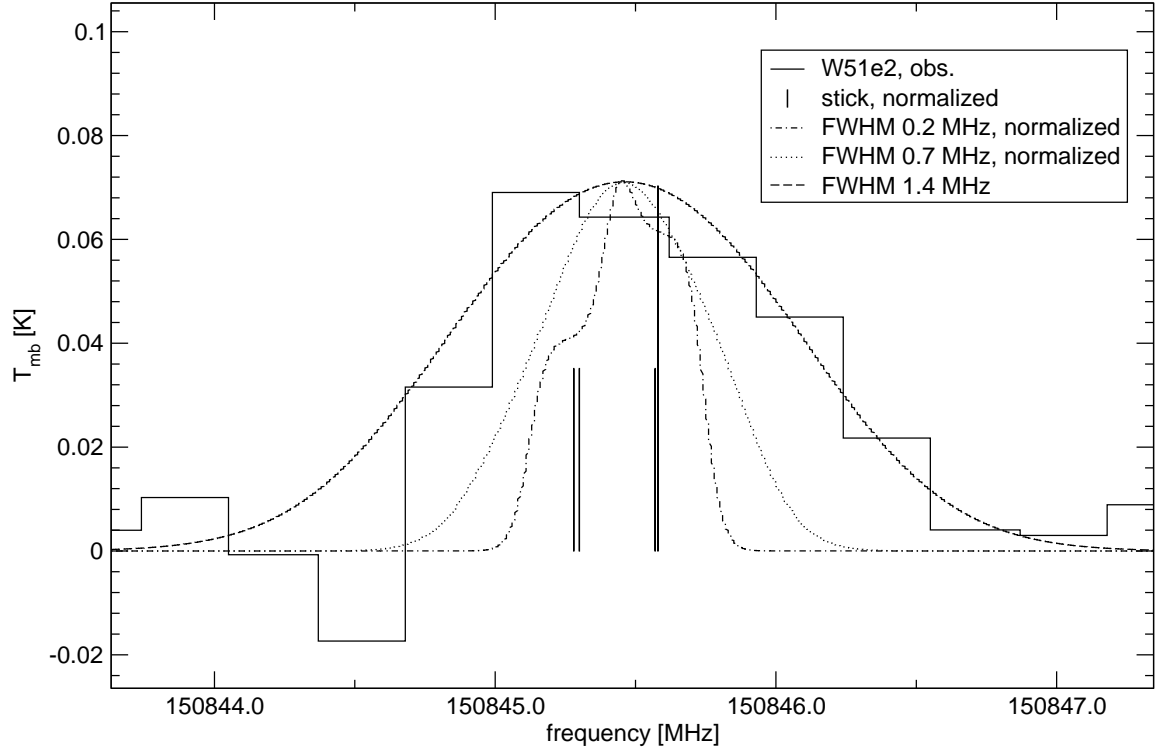}
  \caption{The 20$_{0,20}$-19$_{1,19}$ transition of EME at 150 845.5 MHz. 
           The line profile observed towards W51e2 (bold) is compared to three 
           simulated line profiles (dashed) of 0.2, 0.7, and 1.4 MHz line widths (FWHM). 
           For a line width of 0.2 MHz (FWHM) the splitting due to internal torsional motion 
           of the molecule is partly resolved. The stick diagram indicates the position and
           relative intensities of the split lines based on our recent high resolution
           laboratory measurements \cite{Fuchs-2003-ApJSS-144-277}. For comparison with the observed 
           spectrum all simulated spectra are normalized to have the same peak intensity.
             }
              \label{151GHz-lineshape}
    \end{figure*}

\begin{figure*}[t]
   \centering
   \includegraphics[angle=270,width=9cm]{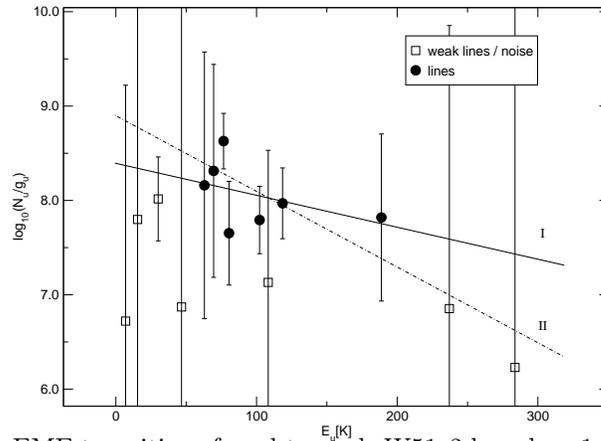}
   \vspace{-0.5cm}\\
   \caption{Boltzmann plot for the EME transitions found towards W51e2 based on 14 observed lines, assuming no  beam dilution. 
            Due to difficulties in assigning the lines to EME two fits have been been 
            performed. I) Fit including all lines marked with $\square$ and $\bullet$ in Table~\ref{EME-trans}. 
            Here T$_{rot}$ = 126$\pm$115~K and log(N$_t$)=14.9$\pm$0.7. 
            II) Fit using only lines with intensities 
            $\ge$ 2$\sigma$ ($\bullet$) resulting in T$_{rot}$ = 54$\pm$38~K and log(N$_t$)=14.3$\pm$1.2. 
            Lines which are less than 2$\sigma$ intense are marked with an $\square$.
            }
              \label{W51e2-EME-boltz}
    \end{figure*}

Several lines at EME transitions have been observed towards W51e2, see 
Fig.~\ref{w51-EME}. 
We looked for strong EME b-type lines and at the same time tried to 
exclude as many blendings with other molecules. 
By using myXCLASS we first estimated a column density of $\sim$10$^{14}$ cm$^{-2}$ and a rotational temperature of roughly 100K.
We checked every frequency band for consistency by comparing a simulated spectrum of EME and other known molecules with 
the observed spectrum.  
Fig.~\ref{151GHz-lineshape} shows an observed line with simulated EME lines superimposed.
The laboratory measurements of the EME line positions are at a precision level unmatched by any astronomical observations. 
The figure illustrates the effect of the linewidth on the 
total observable line shape and shows that the unique EME feature which is due to the 
torsional splitting remains unresolved in the observed spectrum.     
The observed lines\footnote{The 5$_{3,2}$-4$_{2,3}$ transition line at 160.1 GHz from \cite{Charnley-2001-SCAPA-57-685} is not 
listed in Table~\ref{EME-trans} 
and has not been considered in our fits.} 
at transition frequencies of EME are shown in Table~\ref{EME-trans}.
The more stronger lines can be found above 150 GHz whereas many lines at the 1$\sigma$ - 2$\sigma$ level have been observed at lower frequencies. 
Hot cores reveal a great
richness and density in molecular emission lines, 
see Fig.~\ref{245GHz-comp},  
and it is not astonishing that 
77\% of the EME lines in the observed bands are blended by strong lines of other known molecules. 
In our case by far the most blendings occur in the 215 GHz band which is therefore not further detailed in Table~\ref{EME-trans}.
12\% of the blendings are 
at line positions of yet unidentified species. However, from intensity considerations it is clear that these latter lines 
cannot be dominated by EME because in this case EME should have also strong lines at other frequencies where 
we have not observed any strong feature.   
Note that many transition in Table~\ref{EME-trans} have labels with possible blendings. 
By carefully checking the databases and literature but also by fitting 
many other species to our 
spectra we could estimate the abundances of several species which could be potential  competitors to the EME lines.   
To further clarify our detection we made two Boltzmann fits, as shown in 
Fig.~\ref{W51e2-EME-boltz}.   
Here, every line  in Table~\ref{EME-trans} which is marked with an ($\bullet$) or ($\square$) is depicted.
If we use all these 14 lines in the Boltzmann fit (I) we get  $T_{\rm rot}$ = 126$\pm$115~K and 
$N_{\rm t}$= $7.9^{+32}_{-6}\times10^{14}$ cm$^{-2}$ as the beam averaged column density.
Since we have many weak lines (i.e. around 1$\sigma$ level) 
the reliability of the fit can be checked by excluding 
the weaker or doubtful lines marked with ($\square$=noisy-like). 
Fit (II) only includes lines ($\bullet$) with intensities larger than 2$\sigma$.
As a result we obtain  $T_{\rm rot}$ = 54$\pm$38~K and $N_{\rm t}$= $6.3^{+25}_{-5}\times10^{14}$ cm$^{-2}$.  
This fit II should be regarded as more reliable than fit I. 
Furthermore, it has been shown that the excluded weak and noisy lines ($\square$) 
are not in contradiction to fit II. 
Due to the large errors of the weak lines in the Boltzmann plot this is indeed not the case, 
but we would still expect slightly higher intensities at the ($\square$) frequency positions. 

\begin{figure*}[th]
   \centering
   \includegraphics[angle=270,width=9cm]{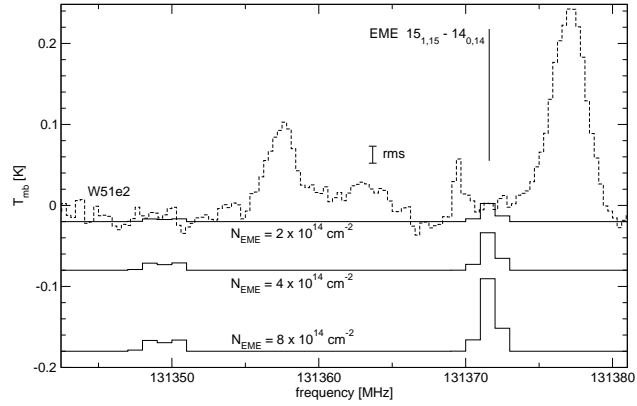}
   \caption{Observed spectrum at W51e2 (top, dashed) and calculated spectra for different column densities of EME at 70K. }
              \label{compare-131GHz-EME-limit-W51e2}
    \end{figure*}

W51e2 is known to have a size $<$ 10'' so that the effect of beam dilution should be considered.
For each fit (I and II) the above mentioned temperatures and abundances of EME can be recalculated by assuming an $5''$ source size.
This results in source averaged column densities and temperatures:
fit I gives $T_{\rm rot}$ = 74$\pm$50~K and $N_{\rm t}$= $2^{+8}_{-1.5}\times10^{14}$ cm$^{-2}$, fit II
$T_{\rm rot}$ = 69$\pm$94~K and $N_{\rm t}$= $2^{+29}_{-1.9}\times10^{14}$ cm$^{-2}$. 
Hence, the column densities and temperatures of the diluted and non-diluted beam yield comparable results within their error limits and we 
will use the source averaged value of fit (II) for further investigations. 
As has already been stated, RTDs cannot necessarily be used as
reliable independent evidence to support the detection of a molecule close to the confusion limit,
as was shown by \cite{Snyder-2005-ApJ-619-914} in the case of glycine.
Our estimated column density is also consistent with observations in less favored frequency regions as can be seen in  
Fig.~\ref{compare-131GHz-EME-limit-W51e2} where an EME line is expected.
For our fitted $2\times10^{14}$ cm$^{-2}$ column density no contradiction within the 1$\sigma$ level can be found, 
but for a $N_{\rm t}=$ 4 - 8 $\times 10^{14}$ cm$^{-2}$ the simulated spectrum is in clear conflict with observation.

It is commonly believed that most complex molecules exist within a small area around the forming star, e.g. around 5'' for W51e2.
On the other hand \cite{Hollis-2001-ApJ-554-L81} have shown that complex molecules such as glycolaldehyde can be found on spatial scales 
of $\geq$ 60$''$ in the case of SgrB2. With our data it is not possible to determine the size of the emitting region but the source 
size has surely an influence on the detected signal strength which we tried 
to estimate using the myXCLASS program. Table~\ref{source-column} lists the result of possible column densities consistent with the 
observations if we assume certain temperatures between 30~K and 150~K.  
\cite{Charnley-2001-SCAPA-57-685} presented a map of a line at an EME transition frequency towards SgrB2 of the order of 20$''$.
If we assume W51e2 to have this same source size the column density would have to be corrected to lower values by 
nearly an order of magnitude. 
More realistic, our best fit value (fit II) assumes an 5'' size of W51e2. 

\begin{figure*}[pth]
   \centering
   \includegraphics[angle=270,width=16cm]{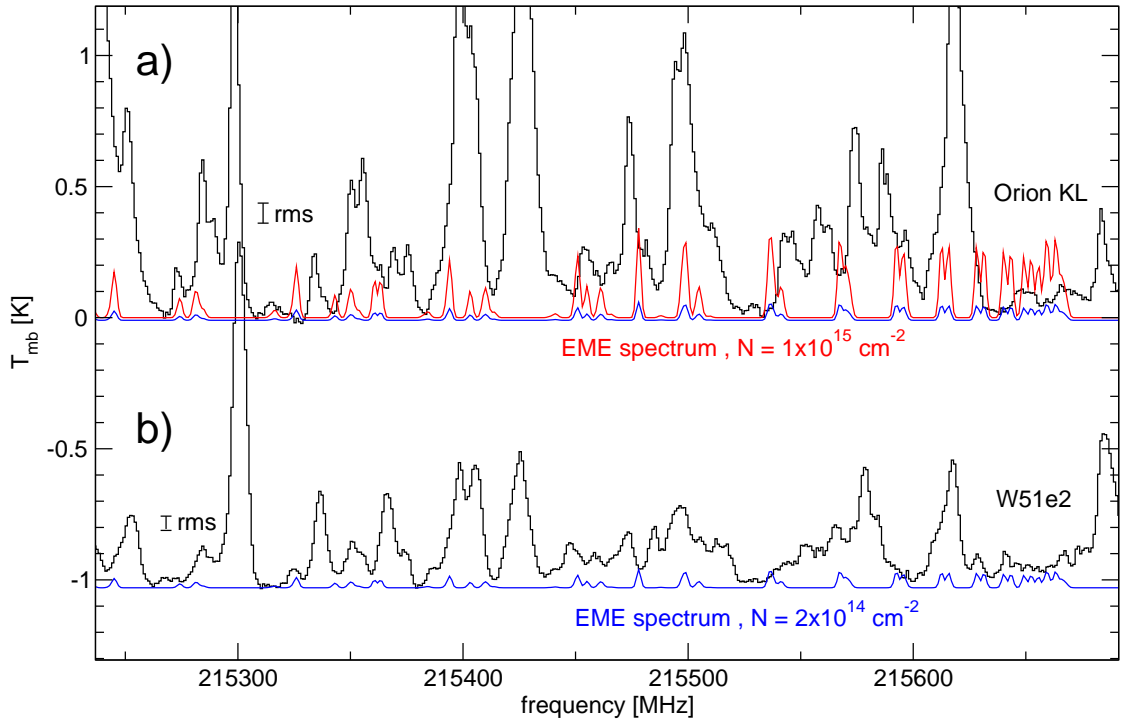}
   \caption{Observed spectra of Orion KL and W51e2 at 215 GHz. 
            This frequency region has the highest line density of all our observed bands but even 
            here gaps can be found where the presence of EME can be checked. 
            Two simulated spectra of EME are plotted, i.e.  
              N=$2\times10^{14}$ cm$^{-2}$ (blue)  and  N=$1\times10^{15}$ cm$^{-2}$ (red) with $T_{\rm rot}$=70~K. 
             The splitting of the lines which can best be seen above 215600 MHz is mainly  
              due to the internal rotation of the methyl groups of EME with an additional small asymmetry contribution.
            a) For Orion KL the previously estimated limits of $10^{14}$ -  $10^{15}$ cm$^{-2}$ by \cite{Charnley-2001-SCAPA-57-685}
               seem questionable as can be seen from the superimposed simulated spectra on the observed spectra.
            b) In the case of W51e2 the simulated spectrum is in agreement with the observation.  
             }
              \label{compare-215GHz}
    \end{figure*}

\begin{figure*}[t]
   \centering
   \includegraphics[angle=270,width=17cm]{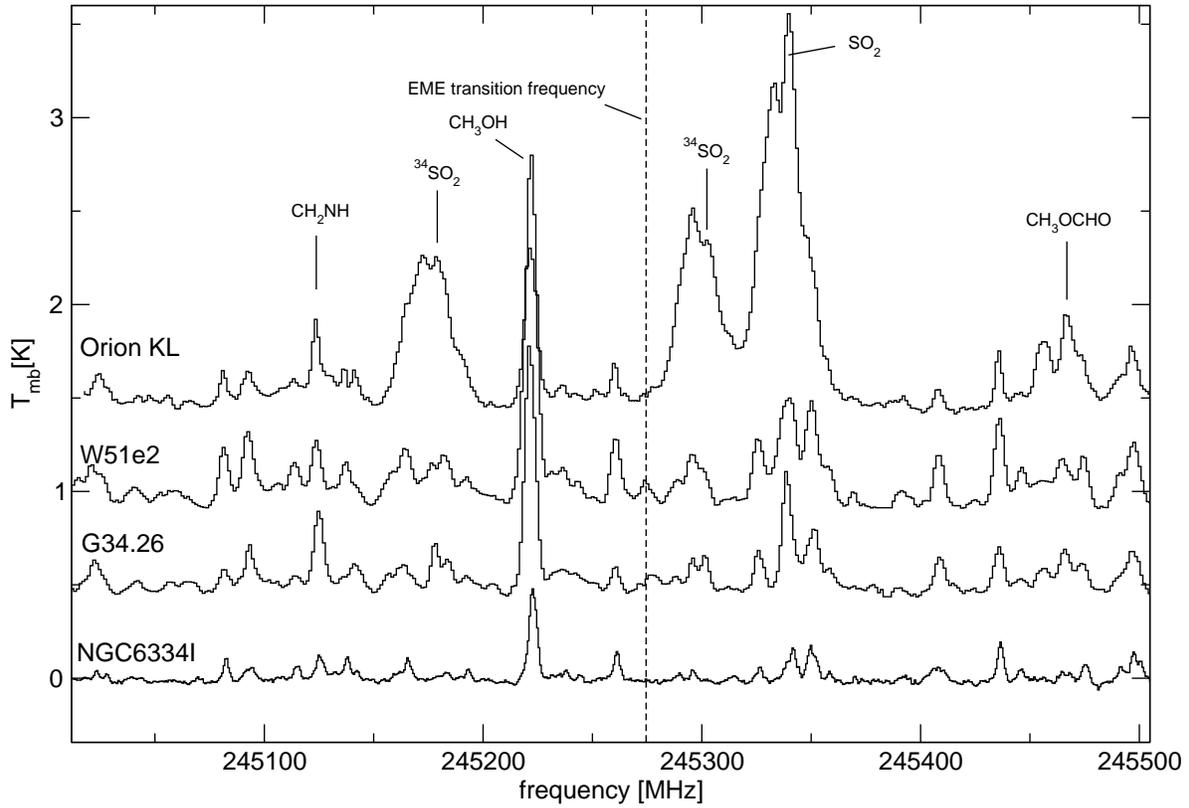}
   \vspace{-0.5cm}\\
   \caption{Spectra of the observed sources at 245 GHz. Only W51e2 reveals a line at an EME transition frequency, whereas all  
            other sources do not show any significant feature above 1$\sigma$ level.  
            However, W51e2, G34.26 and NGC6334I have similar spectra indicating the same chemical composition and equivalent 
            dynamics. 
            }
              \label{245GHz-comp}
    \end{figure*}

Furthermore, it can be seen from Fig.~\ref{compare-215GHz} (see simulated spectrum for N = $1\times$ 10$^{15}$ cm$^{-2}$)
that in some frequency regions strong EME lines, i.e. lines well above 3$\sigma$ level, 
should reveal their internal line splitting. 
Due to the weakness of the observed lines 
no line was detected where the torsional splitting could be resolved so that there is no unique proof that these lines 
indeed belong to EME.

\subsection{The other sources}
  Sample spectra showing all examined sources at 245 GHz are shown in Fig.~\ref{245GHz-comp}.
  At this frequency one of the strongest observed EME candidate lines appears in W51e2.
  However, no line above 1$\sigma$ appears in these spectra for any other examined source. 
  Also for the other frequency bands we get similar results. 
  For example, the well pronounced feature at 150\,845 MHz in W51e2 is only at the 1-2$\sigma$ level in Orion KL and 
  is completely invisible in G34.26 and NGC6334I.
  An upper limit for the column density  of G34.26, NGC6334I and Orion KL has been estimated by the use 
  of myXCLASS and independently by RTDs using integrated noise at EME transition 
  frequencies and assuming an appropriate line width. 
  Fig.~\ref{compare-215GHz} compares the 215 GHz region of W51e2 with Orion KL and also shows simulated spectra 
  of EME for $N$= $2\times10^{14}$ cm$^{-2}$  and $N$= $1\times10^{15}$ cm$^{-2}$ at 70~K (assuming an 5'' source size).
  It is immediately clear that the column densities of these sources must be of the order of $10^{14}$ cm$^{-2}$ or less rather than 
  $10^{15}$ cm$^{-2}$ as estimated by \cite{Charnley-2001-SCAPA-57-685}. 
 
  Our results are summarized in  Table~\ref{abundance-tab}. The previously 
  published line at the EME  5$_{3,2}$-4$_{2,3}$ transition frequency  at 160.1 GHz in Orion KL 
  by \cite{Charnley-2001-SCAPA-57-685} 
  has not been considered in the fit.


\section{Discussion}
\label{disc}

\cite{Charnley-2001-SCAPA-57-685} estimate the column density of EME in Orion KL and SgrB2(N) to 
be between 10$^{14}$ - 10$^{15}$ cm$^{-2}$ and in W51 to be a few times 10$^{14}$ cm$^{-2}$. 
EME column densities at around 10$^{14}$ cm$^{-2}$ correspond to signal strengths which are already close to the 
confusion limit in the HMC spectra. Nevertheless, long integration times and the availability of spectra at many frequencies  
enable us to estimate the column density of Orion KL 
  to be  below 1$\times$ 10$^{14}$ cm$^{-2}$ 
  and those of G34.26 and NGC6334I to be of the same order.   

We can now use our values in Table~\ref{abundance-tab} of W51e2 to draw some conclusions using the 
model of \cite{Charnley-1995-ApJ-448-232}.
According to Fig.~2 and Fig.~3a in their paper, 
which assumes a 10$^{-6}$ fractional abundance $X_{Eol}$ of ethanol
as a starting value for the chemical evolution in their model, the age of W51e2 should be 
either $\ge 10^5$ yr if we use CH$_3$OH or C$_2$H$_5$OH as a probe molecule or 10$^4$ -  10$^5$ if we look at CH$_3$OCH$_3$. 
However, that would imply a fractional abundance between $\le$ 10$^{-10}$ and $\sim 3\times 10^{-9}$ for EME. 
On the other hand, if we use Fig.~4a of 
\cite{Charnley-1995-ApJ-448-232} with ethanol starting at $X_{Eol}$=10$^{-7}$ this translates to 
$X_{\rm EME}$ $\le$ 10$^{-12}$ and $X_{\rm EME}$ $\sim$ $4\times10^{-10}$. 
Assuming C$_2$H$_5$OH to be the best molecule to determine the chemical age of the HMC it appears that 
in both cases, i.e. with 
ethanol starting values or $X_{Eol}$ = 10$^{-6}$ and  $X_{Eol}$ = 10$^{-7}$, the fractional abundance of EME is nearly the same, 
namely $X_{\rm EME} \sim 2 \times 10^{-10}$. 
This value is close to our estimated abundance $5.6 \times 10^{-10}$ for EME in W51e2. 
According to Table~\ref{abundance-tab} and the model by \cite{Charnley-1995-ApJ-448-232} Orion KL and NGC6334F should
be roughly of the same age as W51e2. 
Although we have only upper limits on the EME abundance in these sources it is clear that as a consequence the C$_2$H$_5$OH fractional 
abundance at 0 yr must be 10$^{-7}$ rather than 10$^{-6}$. 
G34.26 is slightly older and SgrB2 is the oldest of the here discussed sources and should have an EME abundance $X_{\rm EME}$ of less than 
10$^{-10}$ (corresponding to less than 10$^{14}$ cm$^{-2}$ column density as we see it for Orion KL, G34.26 and NGC6334I). 
This however is in contradiction to the claimed observations of EME towards SgrB2 by \cite{Charnley-2001-SCAPA-57-685} when 
they estimate an abundance of 10$^{-10}$ - 10$^{-9}$. 
    
\begin{figure*}[t]
   \centering
   \includegraphics[angle=270,width=10cm]{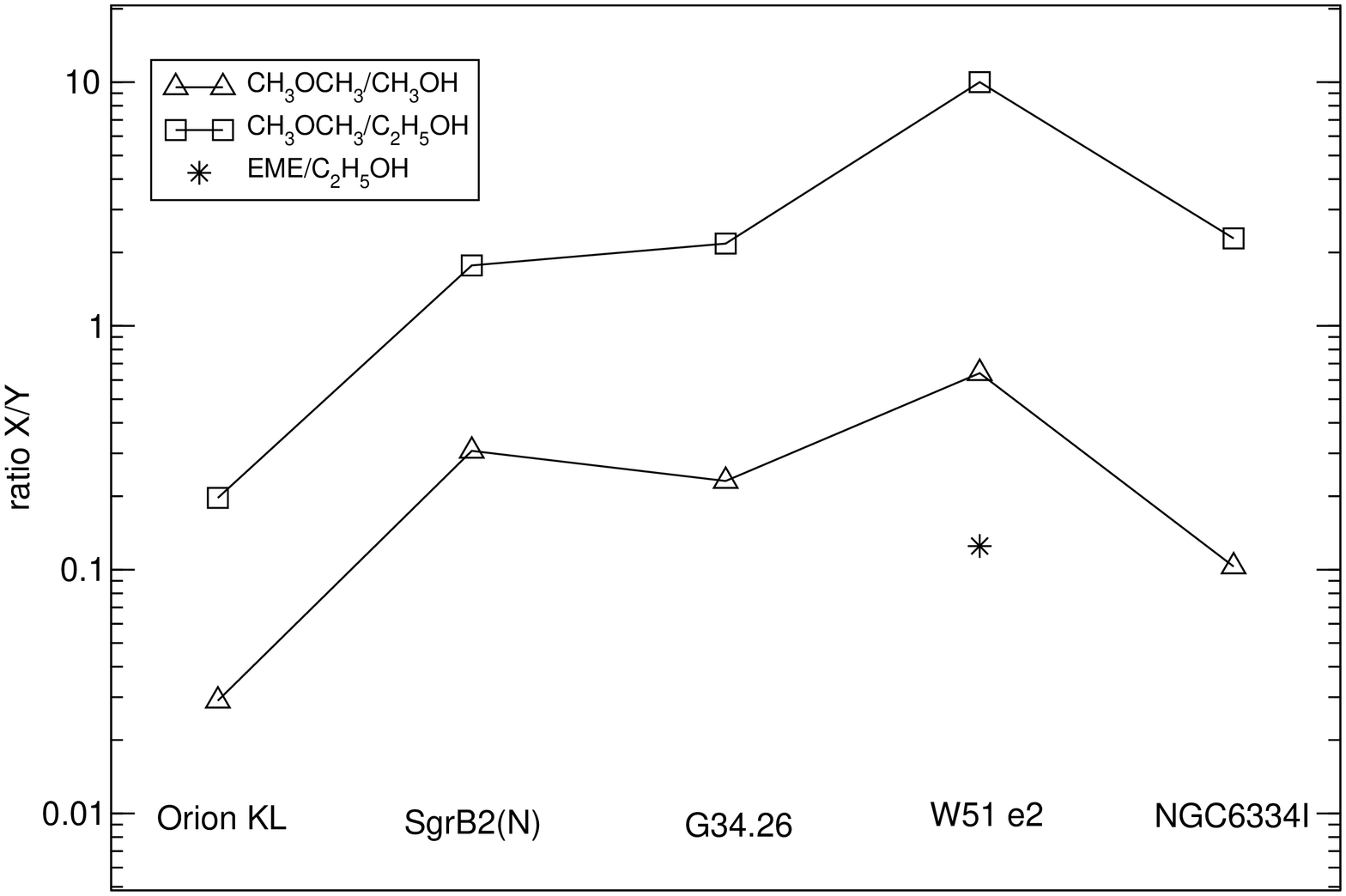}
   \vspace{-1cm}\\
   \caption{Relative abundances of CH$_3$OH, C$_2$H$_5$OH, CH$_3$OCH$_3$ derived from Table~\ref{abundance-tab} by using 
            the values from \cite{Turner-1991-ApJSS-76-617} for Orion KL,  \cite{Turner-1991-ApJSS-76-617} for SgrB2(N), 
            \cite{Thompson-1999-AA-342-809} and  \cite{Ikeda-2002-ApJ-571-560}  
            for G34.26, this work for W51e2 and \cite{Nummelin-1998-AA-337-275} for NGC6334I. 
            The ratio of EME/C$_2$H$_5$OH in W51e2 is 0.13 which is 5 times smaller than the ratio of  CH$_3$OCH$_3$/CH$_3$OH.  
            }
              \label{rel-abund}
    \end{figure*}

Looking at the observed values only, we notice that Orion KL has the highest fractional abundance of 
ethanol relative to H$_2$ followed by NGC6334I, G34.26, W51e2 and then SgrB2(N). 
However, no EME has been detected in  Orion KL. 
The relative abundances of DME with respect to methanol and ethanol are shown in Fig.~\ref{rel-abund}.
Here, 
W51e2 shows a relative high dimethyl ether content relative to methanol and ethanol compared to the other sources, which may be 
a general indication of a higher ether abundance relative to the alcohols.      
Assuming that we indeed have detected EME in W51e2 
the ratio between EME and ethanol is 0.13 which is 5 times less than the ratio of DME and methanol. 
If all sources would have a 0.13 EME/ethanol ratio the theoretical column densities would be around 
2$\times10^{14}$ cm$^{-2}$ 
or higher
\footnote{except for  the SgrB2(N) value by \cite{Cummins-1986-ApJSS-60-819}} 
which at least for G34.26, NGC6334I and Orion KL has 
been shown to be not the case. 
Thus, W51e2 is indeed particularly rich in EME.

We conclude that only the tentative detection of EME by \cite{Charnley-2001-SCAPA-57-685} towards 
W51e2 can be confirmed within the following caveats 
1) the EME Boltzmann fits include a few lines close to the noise level,
2) the obtained temperature of 70 K for EME is lower than that of other 
   observed molecules ($\gtrsim$ 140 K), 
3) no torsional splitting of the EME lines could be observed and also other characteristic spectral fingerprints have not been observed. 
However, there are lines at EME transition frequencies which until now can best be explained by the existence of EME in W51e2.

\begin{acknowledgements}
      We are grateful to 
      the IRAM 30m and SEST 15m staff for their hospitality and assistance. Many thanks to Peter Schilke (MPIfR) 
      for providing us with the myXCLASS program code. Many thanks to Eric Herbst for his invaluable suggestions and discussions.
      This work was supported by the 
      \emph{Deut\-sche For\-schungs\-ge\-mein\-schaft, DFG\/} within SFB 494, IRAM, and ESO.
\end{acknowledgements}


\bibliographystyle{aa}

\clearpage

   
    \begin{table*}
      \caption[]{Astronomical sources}
         \label{astro-sou}
         \begin{center}
         \begin{tabular}{llrcrcc}
            \hline
            \hline
            \noalign{\smallskip}
             Source & \multicolumn{1}{c}{RA} &  \multicolumn{1}{c}{Dec} & Eq. &  \multicolumn{1}{c}{LSR} & telescope \\
                    &                        &                          &     &    (km s$^{-1}$)         & \\         
            \noalign{\smallskip}
            \hline
            \noalign{\smallskip}
        {\bf G34.26}      & 18 53 18.496 &  01 14 58.66 & 2000 & +60.0   & IRAM 30m  \\
    {\bf NGC 6334 I}      & 17 17 32.300 & -35 44 04.00 & 1950 & -8.0    & SEST 15m  \\
      {\bf Orion KL}      & 05 35 14.171 & -05 22 23.10 & 2000 & +8.8    & IRAM 30m  \\
      {\bf W51 e2}        & 19 23 43.959 &  14 30 34.55 & 2000 & +55.0   & IRAM 30m   \\
\noalign{\smallskip}
            \hline
         \end{tabular}
\end{center}
  \end{table*}


\begin{spacing}{1.0}
    \begin{table*}
      {\small{
      \caption[]{Rotational temperatures, column densities and fractional abundances $X$ relative to H$_2$ of CH$_3$OH, 
                 C$_2$H$_5$OH, CH$_3$OCH$_3$ and C$_2$H$_5$OCH$_3$ (EME) from our and other selected works.}
         \label{abundance-tab}
         \begin{center}
         \begin{tabular}{llllrrl}
            \hline
            \hline
            \noalign{\smallskip}
        source &     molecules & & \multicolumn{1}{c}{$T_{\rm rot}$ [K]} &  \multicolumn{1}{c}{$N_{\rm t}$ [cm$^{-2}$]} & $X^{a}$ & Reference \\
             \hline
            \noalign{\smallskip}
   {\bf G34.26} & & & & \\
              & CH$_3$OH          & a-type & 368 $\pm$ 16   & (1.8 $\pm$ 0.2) $\times$10$^{16}$ & 4.5$\times$10$^{-8}$ &\cite{Macdonald-1996-AASS-119-333} \\
              &                   &        & 368 $\pm$ 16   & (1.6 $\pm$ 0.2) $\times$10$^{16}$ & 4$\times$10$^{-8}$ &\cite{Thompson-1999-AA-342-809}$^{b}$ \\
              &                   & b-type & 336 $\pm$ 14   & (1.9 $\pm$ 0.1) $\times$10$^{16}$ & 4.8$\times$10$^{-8}$ &\cite{Macdonald-1996-AASS-119-333} \\
              &                   & a+b type & 96  $\pm$ 17 & (2.6 $\pm$ 1.4) $\times$10$^{16}$  & 6.5$\times$10$^{-8}$ &\cite{Ikeda-2002-ApJ-571-560} \\
              &                   & a-type   & 140 $\pm$ 14   &  2.2$^{+4.7}_{-0.4}$ $\times$10$^{16}$ & 5.5$\times$10$^{-8}$ &this work \\ 
              & C$_2$H$_5$OH      &        & 94 $\pm$ 27    & (3.5 $\pm$ 2.0) $\times$10$^{15}$ & 8.8$\times$10$^{-9}$ &\cite{Macdonald-1996-AASS-119-333} \\
              &                   &        & 75 $\pm$ 37    & (1.7  $\pm$ 2.1) $\times$10$^{15}$ & 4.3$\times$10$^{-9}$ &\cite{Ikeda-2002-ApJ-571-560} \\
              &                   &        & 152 $\pm$ 66   & 6.0$^{+8.4}_{-3.5}$ $\times$10$^{14}$ & 1.5$\times$10$^{-9}$ & this work \\
              & CH$_3$OCH$_3$     &        & 137 $\pm$ 32   &  (3.7 $\pm$ 1.8) $\times$10$^{15}$ & 9.3$\times$10$^{-9}$ &\cite{Ikeda-2002-ApJ-571-560} \\
              &                   &        & 146 $\pm$ 38   &  3.5$^{+1.8}_{-1.2}$ $\times$10$^{15}$ & 8.8$\times$10$^{-9}$ & this work \\
              & C$_2$H$_5$OCH$_3$  &       & 100 - 150      & $<$ 7 $\times$10$^{13}$ & $<$1.8$\times$10$^{-10}$ & this work\\
  {\bf NGC 6334 I}  & & & &\\
              & CH$_3$OH          & & 79$^{+21}_{-8}$          & 3.1$^{+0.8}_{-1.0}$  $\times$10$^{16}$ & 1.6$\times$10$^{-7}$ &\cite{Nummelin-1998-AA-337-275} \\
              &                   & & 66  $\pm$ 11             & 1.2$^{+1.9}_{-1.2}$ $\times$ 10$^{16}$    & 6.0$\times$10$^{-8}$  & this work \\
              & C$_2$H$_5$OH      & & 94$^{+44}_{-20}$         & 1.4$^{+0.5}_{-0.3}$ $\times$10$^{15}$ &  7.0$\times$10$^{-9}$ &\cite{Nummelin-1998-AA-337-275} \\
              &                   & & 120 $\pm$ 44             &  4.0$^{+1.2}_{-0.9}$ $\times$ 10$^{14}$   &  2.0$\times$10$^{-9}$     & this work \\
              & CH$_3$OCH$_3$     & & 20/440                   & 3.5$\times$10$^{15}$ / 1.3$\times$10$^{16}$ & 4.1$\times$10$^{-8}$   & \cite{Nummelin-1998-AA-337-275} \\
              &                    & & 119 $\pm$ 39            &  2.3$^{+1.8}_{-1.0}$ $\times$ 10$^{15}$   &  1.2$\times$10$^{-8}$    & this work \\
              & C$_2$H$_5$OCH$_3$  & & 50 - 150                & $<$ 8 $\times$ 10$^{13}$  &  $<$4$\times$10$^{-10}$   & this work \\
   {\bf Orion KL}  & & & &\\
              & CH$_3$OH          & &  120 - 140               & 5 $\times$ 10$^{16}$ & 5$\times$10$^{-7}$  & \cite{Sutton-1985-ApJSS-58-341} \\
              &                   & &  140                     &  4.5 $\times$10$^{16}$    &  4.5$\times$10$^{-7}$ &\cite{Turner-1991-ApJSS-76-617} (Onsala)\\
              &                   & &  192$^{+18}_{-16}$       & (1.6$\pm$0.2) $\times$10$^{16}$ &  1.6$\times$10$^{-7}$ &\cite{Turner-1991-ApJSS-76-617} (NRAO)\\
              &                   & & $>$ 200                  & 2.5 - 30 $\times$10$^{16}$ &  1.6$\times$10$^{-6}$ &\cite{Sutton-1995-ApJSS-97-455}  \\
              &                   & & 120-200                  & 4-8 $\times$ 10$^{16}$      &  6$\times$10$^{-7}$ &this work \\
              & C$_2$H$_5$OH      & & 216$^{+3600}_{-105}$     & 6.6$^{+4.1}_{-2.5}$ $\times$ 10$^{15}$ &  6.6$\times$10$^{-8}$ &\cite{Turner-1991-ApJSS-76-617} \\
              &                   & & 289 $\pm$ 330          &  4.8$^{+35}_{-4.2}$ $\times$ 10$^{14}$  & 4.8$\times$10$^{-9}$ &this work$^{c}$  \\
              & CH$_3$OCH$_3$     & & 75                       &  2.5 $\times$ 10$^{15}$  &  2.5$\times$10$^{-8}$ &\cite{Johansson-1984-AA-130-227} \\
              &                   & & 63                       &  3.0 $\times$ 10$^{15}$  &  3.0$\times$10$^{-8}$ &\cite{Blake-1986-ApJSS-60-357} \\
              &                   & & 91$^{+34}_{-19}$       & 1.3$^{+0.3}_{-0.1}$ $\times$10$^{15}$ &  1.3$\times$10$^{-8}$ &\cite{Turner-1991-ApJSS-76-617} \\
              &                   & & 125                      & $\sim$ 3 $\times$10$^{15}$ &   3.0$\times$10$^{-8}$ &\cite{Sutton-1995-ApJSS-97-455} \\
              &                   & & 197.4$^{+165}_{-62}$ &  1.1$^{+2.5}_{-0.6}$ $\times$10$^{16}$ &  1.1$\times$10$^{-7}$ &\cite{Lee-2002-JKAS-35-187} \\
              &                   & & 126$\pm$43               &  4.6$^{+3.1}_{-1.8}$ $\times$ 10$^{15}$  &  4.6$\times$10$^{-8}$ & this work  \\
              & C$_2$H$_5$OCH$_3$ & & 100-150                  & $<$7 $\times$ 10$^{13}$   &  $<$7$\times$10$^{-10}$ & this work  \\
   {\bf SgrB2(N)}  & & & &\\
              & CH$_3$OH$^{d}$ & I  & 49  $\pm$ 1         & (2.8 $\pm$ 0.07) $\times$10$^{16}$   &   5.6$\times$10$^{-9}$   & \cite{Cummins-1986-ApJSS-60-819}\\
              &                &    & 26$^{+138}_{-12}$ & 8.1$^{+8}_{-4}$ $\times$10$^{15}$   &  1.6$\times$10$^{-9}$  & \cite{Turner-1991-ApJSS-76-617} \\
              &                & II & 120 $\pm$ 20        & (3.5 $\pm$ 0.5) $\times$10$^{15}$   &  7$\times$10$^{-10}$    & \cite{Cummins-1986-ApJSS-60-819}\\
              &                &    & 204$^{+392}_{-81}$ & 6.9$^{+5.6}_{-3.1}$ $\times$10$^{15}$ &  1.4$\times$10$^{-9}$ & \cite{Turner-1991-ApJSS-76-617} \\
              & C$_2$H$_5$OH   & & 14                  & 5.3 $\times$10$^{14}$                 &  1.1$\times$10$^{-10}$ &\cite{Cummins-1986-ApJSS-60-819}\\
   & &                             & 36                & 4.8$^{+4.5}_{-2.4}$ $\times$10$^{15}$ &  9.6$\times$10$^{-10}$ &\cite{Turner-1991-ApJSS-76-617} \\
   &           CH$_3$OCH$_3$  &   & 24                  & 5 $\times$10$^{14}$                   &  1$\times$10$^{-10}$ &\cite{Cummins-1986-ApJSS-60-819}\\
   & &                            & 80$^{+351}_{-37}$ & 2.3$^{+0.9}_{-0.4}$ $\times$10$^{15}$ &  4.6$\times$10$^{-10}$ &\cite{Turner-1991-ApJSS-76-617} $^{e}$\\
  {\bf W51e2} & & & & \\
              & CH$_3$OH,      &     & 208         & 1.0 $\times$10$^{17}$   &    2.7$\times$10$^{-7}$ & \cite{Ikeda-2002-ApJ-571-560} \\   
              &                & a-type  & 101$\pm$4   & 2.5$^{+0.5}_{-0.3}$ $\times$10$^{16}$  &    6.9$\times$10$^{-8}$            & this work \\
              & C$_2$H$_5$OH   &     & 169$\pm$28  & (3.1 $\pm$ 1.1)   $\times$10$^{15}$  &   8.6$\times$10$^{-9}$ &  \cite{Ikeda-2002-ApJ-571-560} \\
              &                &     & 143$\pm$20  & 1.6$^{+0.4}_{-0.3}$ $\times$10$^{15}$ &  4.4$\times$10$^{-9}$  & this work \\
              & CH$_3$OCH$_3$  &     & 145$\pm$8   & 1.6$^{+0.4}_{-0.2}$ $\times$10$^{16}$  & 4.4$\times$10$^{-8}$ & this work\\
              & C$_2$H$_5$OCH$_3$ &  & 69$\pm$94 & 2 $^{+29}_{-1.9}$ $\times$10$^{14}$      &  5.6$\times$10$^{-10}$  & this work\\
            \noalign{\smallskip}
            \hline
            \noalign{\smallskip}  
\end{tabular}
\end{center}  
\begin{list}{}{}
\item[$^{\mathrm{a}}$]   We used N(H$_2$)(G34.26)  = 4$\times$10$^{23}$ cm$^{-2}$,
                                 N(H$_2$)(NGC6334I)= 2$\times$10$^{23}$ cm$^{-2}$,
                                 N(H$_2$)(Orion KL)= 1$\times$10$^{23}$ cm$^{-2}$,
                                 N(H$_2$)(SgrB2(N))= 5$\times$10$^{24}$ cm$^{-2}$ and
                                 N(H$_2$)(W51e2)   = 3.6$\times$10$^{23}$ cm$^{-2}$,
                         see section~\ref{analysis}.
\item[$^{\mathrm{b}}$]   Observed at center of G34.26, as opposed to the halo.
\item[$^{\mathrm{c}}$]  The Onsala and CIT survey (see \cite{Turner-1991-ApJSS-76-617}) did not find any ethanol towards Orion KL 
                        and estimated 
                        an upper limit between 1.5 and 4$\times$10$^{14}$ cm$^{-2}$. 
\item[$^{\mathrm{d}}$] \cite{Cummins-1986-ApJSS-60-819} and \cite{Turner-1991-ApJSS-76-617} used two rotational temperature diagrams 
                       (I and II) to fit the observed CH$_3$OH lines. \cite{Cummins-1986-ApJSS-60-819} uses the transitions with E$_u$ 
                       below 70 K for the fit I whereas  \cite{Turner-1991-ApJSS-76-617} splits the two components I and II at  
                       $E_{\rm u}$ = 40 K. 
\item[$^{\mathrm{e}}$]  \cite{Turner-1991-ApJSS-76-617} gives two fits of $T_{\rm rot}$ and $N_{\rm t}$, the values used here correspond to the most 
                        reliable fit.
\end{list}
}}
  \end{table*}

   
    \begin{table*}
{\small{
      \caption[]{EME transitions observed in W51e2}
         \label{EME-trans}
         \begin{center}
         \begin{tabular}{c...c..ll}
            \hline
            \hline
            \noalign{\smallskip}
Transition & \multicolumn{1}{c}{Observed}  &  \multicolumn{1}{c}{Theor.}     &  \multicolumn{1}{c}{Obs-calc} & line &  \multicolumn{1}{c}{E$_{upper}$} &  \multicolumn{1}{c}{$\int T_{mb} d\nu$} \\
          &  \multicolumn{1}{c}{frequency$^a$} &  \multicolumn{1}{c}{frequency$^c$}  &          &   components     &             &                    \\
          &  \multicolumn{1}{c}{(MHz)}     &   \multicolumn{1}{c}{(MHz)}     &   \multicolumn{1}{c}{(MHz)}   &        &  \multicolumn{1}{c}{(K)}         &  \multicolumn{1}{c}{(K km s$^{-1}$)}    \\
            \noalign{\smallskip}
            \hline
            \noalign{\smallskip}
 11$_{2,10}$-11$_{1,11}$ &  80\,882.3(5)^b    &  80\,882.7(10)^d  &  -0.4 & 5 & 30.1  &  0.09(4) & $\square$  & n     \\
 24$_{1,23}$-24$_{0,24}$ &  81\,198.4(1)    &  81\,198.7(5)   &  -0.3 & 5 & 118.5 &  0.16(6)         & $\bullet$  & c$_1$ \\
\\ 
 35$_{3,32}$-35$_{2,33}$  &  91\,439.6(5)   &  91\,440.4(10)  & -0.8  & 5 & 255.2 &  1.0(44)   &               & b     \\
 36$_{3,33}$-36$_{2,34}$  &      -          &  91\,475.6(9)   &       & 5 & 269.3 &            &               & b     \\ 
 34$_{2,32}$-34$_{1,33}$  &  91\,630.1^e    &  91\,630.8(5)   & -0.7  & 5 & 237.1 &  0.03(9)   & $\square$     & c$_2$, n \\ 
 34$_{3,31}$-34$_{2,32}$  &      -          &  91\,688.3(10)  &  -    & 5 & 241.5 &   -        &               & n     \\ 
 37$_{3,34}$-37$_{2,35}$  &  91\,812.7(4)   &  91\,812.6(9)   & 0.1   & 5 & 283.7 &  0.1(5)    & $\square$     & n     \\
\\
 35$_{2,33}$-35$_{1,34}$ &      -           &  96\,207.0(9)   &  -    & 5 & 250.8 &  -       &               & b      \\ 
 29$_{3,26}$-29$_{2,27}$ &  96\,390.8(3)    &  96\,391.4(12)  & -0.6  & 5 & 179.1 & 1.85(37) &               & b      \\  
 3$_{2,1}$- 2$_{1,2}$   &  96\,464.1^e     &  96\,464.3(6)   & -0.2  & 5 & 6.9   & 0.002(5)  & $\square$     & n      \\
\\
 22$_{3,19}$-22$_{2,20}$ & 107\,654.8^e  & 107\,656.8(14)  & -2.0  & 5 & 108.3 & 0.05(7)     & $\square$ &  n     \\ 
\\
 7$_{2,5}$-6$_{1,6}$     & 131\,350.2(10)  & 131\,350.7(8) & -0.5   & 5 & 15.4  &  0.02(6)    & $\square$ & c$_3$, n \\ 
 15$_{1,15}$-14$_{0,14}$ & 131\,372.68(80) & 131\,372.8(5) &  -0.12 & 5 & 46.7  &  0.02(10)   & $\square$ & c$_4$, n \\ 
\\
 36$_{1,35}$-36$_{0,36}$ &      -           & 150\,502.4(19) &      & 5 & 260.0 6 &   -      &               & b    \\ 
 34$_{4,30}$-34$_{3,31}$ & 150\,661.3(6)    & 150\,662.9(16) & -1.6 & 5 & 248.7 &  0.12(2)   &               & b    \\
 13$_{6,x}$-14$_{5,y}$   & 150\,794.1^e   & 150\,794.1(18) &  0.0  &  10   & 76.6  &  0.17(0.05) & $\bullet$ & c$_5$ \\ 
 20$_{0,20}$-19$_{1,19}$ & 150\,845.52(30)  & 150\,845.5(2) &  0.02 & 5 & 80.4  &  0.20(11)      & $\bullet$ & c$_6$ \\
\\
\, [...]$^f$ &                 &                &      &   &       &           &            & b \\
 19$_{5,x}$-19$_{4,15}$ & 215\,325.4^e    & 215\,326.3(13) & -0.9 & 6 & 102.2 & 0.28(10)  & $\bullet$  &   \\
\, [...]$^f$              &                 &                &      &   &       &           &            & b \\
\\
 28$_{0,28}$-27$_{1,27}$ & 217\,939.6(3) & 217\,940.7(1)  & -1.1  & 5 & 154.7 & 3.66(19)  &            & b \\ 
\\
 16$_{3,14}$-15$_{2,13}$ & 245\,106.4(32) & 245\,104.7(29) & 1.43  & 5 & 62.9  &  0.46(65)   & $\bullet$  & c$_7$ \\ 
 31$_{1,31}$-30$_{0,30}$ & 245\,275.0(13) & 245\,274.1(1)  & 0.86  & 5 & 188.8 &  0.87(77)   & $\bullet$  &       \\ 
\\
 17$_{3,15}$-16$_{2,14}$ & 252\,190.7(11) & 252\,189.4(29) & 1.3 & 5 & 69.5  &  0.70(79)        & $\bullet$  & c$_8$  \\ 
 28$_{2,27}$-27$_{1,26}$  & 253\,309.3(3) & 253\,308.2(12) &  1.09 & 5 & 161.0 &   16.01(126)   &            & b  \\ 
\noalign{\smallskip}
            \hline
         \end{tabular}
\end{center}
\begin{list}{}{}
\item[$^{\mathrm{a}}$] Rest frequencies (assuming a source LSR velocity of +55 km s$^{-1}$)
                       in W51e2.
\item[$^{\mathrm{b}}$] The numbers in parenthesis are the estimated uncertainties (1$\sigma$). 
\item[$^{\mathrm{c}}$] Frequencies are taken from  \cite{Fuchs-2003-ApJSS-144-277}.
\item[$^{\mathrm{d}}$] The deviation from the 
mean frequency of the transitions to its extreme values, e.g. due to rotational and internal rotational splitting the transition  
11$_{2,10}$-11$_{1,11}$ is split in 5 transitions which occur between 80\,881.7 - 80\,883.6 GHz, thus the frequency denoted in the 
table is 80\,882.7(10).
\item[$^{\mathrm{e}}$] Manually determined (or in case of noise the closest peak). Here the transition 13$_{6,x}$-15$_{5,y}$ are with  
                       k$_c$= 7-9, 7-10, 8-9, 8-10 and the x in 19$_{5,x}$-19$_{4,15}$ denotes either 14 or 15.
\item[$^{\mathrm{f}}$] Region with many blends of EME with other (dominating) molecules. 
\item[$\bullet$] Lines included in the Boltzmann fit II, see Fig.~\ref{W51e2-EME-boltz}.
\item[$\square$] Lines included in the Boltzmann fit I and II, see Fig.~\ref{W51e2-EME-boltz}.
\item[n]  ''Noisy'' line, i.e. a line which has either less than 2$\sigma$ intensity or where the intensity is difficult to 
          obtain due to baseline variations.  
\item[b]  Denotes a blend with an other molecular line where the other line contributes most to the total intensity. 
\item[c$_i$] Denotes a contribution of another molecule to the total intensity where the blended line is estimated to contribute less than 20\% 
             (if not otherwise mentioned) compared to EME. Assuming T$_{rot}$ $\sim$ 100~K: \\
             c$_1$ C$_2$H$_5$OOCH with $N\le$5$\cdot10^{15}$ cm$^{-2}$ (could be up to 50\%),
             c$_2$ HOONO$_2$ with $N\le$5$\cdot10^{14}$ cm$^{-2}$,
             c$_3$ c-H$_2$C$_3$O with $N\le$2$\cdot10^{12}$ cm$^{-2}$,
             c$_4$ C$_2$H$_3$NH$_2$ with $N\le$8$\cdot10^{15}$ cm$^{-2}$,
             c$_5$ C$_3$H$_7$CN with $N\le$2$\cdot10^{14}$ cm$^{-2}$,
             c$_6$ C$_2$H$_5$C$^{15}$N with $N\le$3$\cdot10^{13}$ cm$^{-2}$ based on our 
                     observation that C$_2$H$_5$CN has $N\sim3\cdot$10$^{14}$ cm$^{-2}$,
             c$_7$ CH$_2$(OH)CHO with $N\le$8$\cdot10^{14}$ cm$^{-2}$,
             c$_8$ a-H$_2$C=CHOH with $N\le$2$\cdot10^{14}$ cm$^{-2}$.
\\
\end{list}
}}
  \end{table*}
\end{spacing}
   
    \begin{table*}
      \caption[]{Estimated column density for various source sizes and rotational temperatures for W51e2.}
         \label{source-column}
         \begin{center}
         \begin{tabular}{ll|l}
            \hline
            \hline
            \noalign{\smallskip}
              \multicolumn{1}{c}{source size} &  \multicolumn{1}{c|}{$T_{\rm rot}$} & \multicolumn{1}{c}{column density} \\
              \multicolumn{1}{c}{[arc sec]}   &  \multicolumn{1}{c|}{[K]}       & \multicolumn{1}{c}{[cm$^{-2}$]}    \\
            \noalign{\smallskip}
            \hline
            \noalign{\smallskip}
               20''     &   30        &   $\sim 5 \times 10^{13}$ \\  
                        &   50        &   $\sim 4 \times 10^{13}$ \\ 
                        &   70        &   $\sim 4 \times 10^{13}$ \\   
               10''     &   50        &   $\sim 8 \times 10^{13}$ \\ 
                        &   70        &   $\sim 9 \times 10^{13}$ \\   
                        &   150       &   $1.3 \times 10^{14}$ \\   
                5''     &   70        &   $2.0 \times 10^{14}$ \\   
                        &   150       &   $4.5 \times 10^{14}$ \\   
\noalign{\smallskip}
            \hline
         \end{tabular}
\end{center}
  \end{table*}

   
    \begin{table*}
      \caption[]{Observational parameters}
         \label{obspar}
         \begin{center}
         \begin{tabular}{cc.c.ccc.}
            \hline
            \hline
            \noalign{\smallskip}
    Receiver  & $T_{\rm sys}^{(a)}$ & \multicolumn{1}{c}{$\tau$} & Tot. integr.  &  \multicolumn{1}{c}{$\rho^{(b)}$} & Telescope & 1$\sigma$ (rms) & Band  &  \multicolumn{1}{c}{Frequency} \\
    frequency &           &      & time (on+off) &            & beam HPBW & noise level    & width &  \multicolumn{1}{c}{resolution} \\
     (MHz)    &  (K)      &      & (min)         &            & ('')      & (K as $T_{\rm A}^*$) & (MHz) &  \multicolumn{1}{c}{(MHz)} \\
            \noalign{\smallskip}
            \hline
            \noalign{\smallskip}
\multicolumn{1}{l}{\bf G34.26} & & & & & & & &  \\
                 81 041    &  144      & 0.03   &  85 & 0.79 & 30 & 0.012 &  490 & 0.3125 \\
                 91 626    &  122      & 0.03   & 115 & 0.77 & 26 & 0.010 &  490 &  0.3125 \\
                 96 371    &  137      & 0.02   & 145 & 0.77 & 25 & 0.014 &  490 & 0.3125 \\
                131 372    &  200      & 0.06   & 115 & 0.71 & 18 & 0.018 &  490 & 0.3125 \\
                150 750    &  307      & 0.04   & 220 & 0.68 & 16 & 0.017 &  490 & 0.3125 \\
                215 500    &  388      & 0.22   & 196 & 0.56 & 11 & 0.032 & 1020 &  1.0 \\
                217 900    &  350      & 0.14   & 147 & 0.55 & 11 & 0.023 & 1020 &  1.0 \\
                245 260    &  511      & 1.90   & 160 & 0.50 & 10 & 0.034 &  510 & 1.25 \\
                252 227    &  848      & 0.37   &  85 & 0.48 & 10 & 0.059 &  510 & 1.25 \\
\multicolumn{1}{l}{\bf NGC 6334 I} & & & & & & & &  \\
                81 198     &  327      & 0.41   & 38  & 0.76 & 59 & 0.009 &  994 & 0.6911 \\
                91 450     &  195      & 0.26   & 37  & 0.74 & 55 & 0.007 &  994 & 0.6911 \\
                91 630.7   &  185      &   -    & 87  & 0.74 & 55 & 0.005 &  994 & 0.6911 \\
                91 811.7   &  212      & 0.06   & 57  & 0.74 & 55 & 0.007 &  994 & 0.6911 \\
                96 380     &  252      & 0.21   & 55  & 0.73 & 53 & 0.009 &  994 & 0.6914 \\
               150 662     &  263      & 0.34   & 95  & 0.65 & 33 & 0.012 &  994 & 0.6989 \\
               245 105     &  513      & 0.96   & 77  & 0.47 & 22 & 0.024 &  994 & 0.6989 \\
               245 274     &  534      & 0.50   & 127 & 0.47 & 22 & 0.017 &  994 & 0.6989 \\
\multicolumn{2}{l}{\bf Orion KL} & & & & & & &  \\
               81 041      &  138      & 0.03   & 125 & 0.79 & 30 & 0.012 &  490 &  0.3125 \\ 
               91 626      &  182      & 0.07   &  95 & 0.77 & 26 & 0.014 &  490 &  0.3125 \\ 
              150 750      &  316      & 0.14   & 220 & 0.68 & 16 & 0.029 &  490 &  0.3125 \\ 
              215 500      &  539      & 0.33   & 160 & 0.56 & 11 & 0.078 & 1020 &  1.0    \\
              245 275      &  740      & 0.39   & 190 & 0.50 & 10 & 0.053 &  510 &  1.25   \\
\multicolumn{2}{l}{\bf W51e2} & & & & & & &  \\
               81 041      &  113      & 0.04   &  95 & 0.79 & 30 & 0.012 &  490 &  0.3125 \\  
               91 626      &  116      & -      &  60 & 0.77 & 26 & 0.012 &  490 &  0.3125 \\  
               96 371      &  119      & 0.07   &  95 & 0.77 & 25 & 0.013 &  490 &  0.3125 \\  
              107 650      &  223      & 0.14   &  52 & 0.75 & 22 & 0.018 &  490 &  0.3125 \\  
              131 372      &  179      & -      &  60 & 0.71 & 18 & 0.019 &  490 &  0.3125 \\  
              150 750      &  249      & 0.26   & 215 & 0.68 & 16 & 0.020 &  490 &  0.3125 \\  
              215 500      &  221      & 0.415  & 195 & 0.56 & 11 & 0.063 & 1020 &  1.0    \\
              217 900      &  301      & 0.634  & 112 & 0.55 & 11 & 0.044 & 1020 &  1.0    \\
              245 260      &  371      & 0.75   &  75 & 0.50 & 10 & 0.051 &  510 &  1.25   \\
              252 227      &  384      & 0.51   &  95 & 0.48 & 10 & 0.062 &  510 &  1.25   \\
              253 300      &  330      & -      &  85 & 0.48 & 10 & 0.045 &  510 &  1.25   \\
\noalign{\smallskip}
            \hline
         \end{tabular}
\\ $^{(a)}$ average system temperature
\\ $^{(b)}$ $T^*_A = \rho T_{mb}$
\end{center}
  \end{table*}

\clearpage



%


%

\end{document}